\documentstyle[12pt,a4,subeqnarray]{article}

\newcommand{\D}{\mbox{d}}
\newcommand{\HRG}{H^{\mbox{\scriptsize RG}}}
\newcommand{\al}[1]{\alpha_{#1}}
\newcommand{\fracp}[2]{\frac{\partial #1}{\partial #2}}

\begin{document}

\begin{flushright}
DPNU-98-18\\ 
April 1998
\end{flushright}

\begin{center}
	{\LARGE Renormalization group equations and integrability in
	Hamiltonian systems }\\
\vspace*{1em}
{\large Yoshiyuki Y. YAMAGUCHI{$^\ast$}
	\footnote{E-mail address: yamaguchi@allegro.phys.nagoya-u.ac.jp}
  and Yasusada NAMBU{$^{\ast\ast}$}
	\footnote{E-mail address: nambu@allegro.phys.nagoya-u.ac.jp} \\
\vspace*{0.5em}	
  {$\ast$:} The general research organization of science and
	  engineering, Ritsumeikan Univ.,\\ 
	  Kusatsu, Shiga, 525-8577, Japan\\ 
  {$\ast\ast$:} Department of Physics, School of Science, \\
	  Nagoya  University,\\ 
	  Nagoya, 464-8602, Japan}\\
\vspace*{1em}
\end{center}
\vspace*{1em}

\begin{abstract}
We investigate Hamiltonian systems with two degrees of freedom
by using renormalization group method. 
We show that the original Hamiltonian systems and
the renormalization group equations are integrable
if the renormalization group equations are Hamiltonian systems
up to the second leading order of a small parameter.
\end{abstract}

Many systems in the nature are described by 
Hamiltonian equations of motion,
and we are interested in dynamical properties of the systems.
Although behavior of systems is revealed
by solving equations of motion,
we cannot generally obtain exact solutions to Hamiltonian systems
with more than one degree of freedom because of lack of integrals. 
We therefore solve equations of motion by using perturbation techniques.
Naive perturbation often fails to give global solutions
since it yields secular terms,
and hence improved perturbation methods have been developed,
for instance,
averaging method, multiple scale method, matched asymptotic expansions
and canonical perturbation theory\cite{hori-66,deprit-69}.

Recently, renormalization group method \cite{chen-96,kunihiro-95}
is proposed as one of the most powerful tools
to construct approximate solutions
since it unifies many of the perturbation techniques listed above.
This method reduces equations of motion to amplitude equations
called renormalization group equations (RGEs) by ignoring fast motion.
The reduced equations, RGEs, must reflect features of 
original systems,
and we expect that dynamical properties of the original systems
are obtained from RGEs.
In this article, we consider two types of perturbed 
Hamiltonian systems with two degrees of freedom.
We investigate symplectic properties, integrability 
and integrals of RGEs,
which are related to properties in the original systems.
We first present our main result: 
\vspace{1ex} \\
\noindent {\bf Theorem }{\it 
  Let Hamiltonian be represented as follows:
  \begin{eqnarray}
    \label{Hamiltonian}
    H(q_1,q_2,p_1,p_2) &=& H_0(q_1,q_2,p_1,p_2) + \epsilon V_1(q_1,q_2),\\
    H_0(q_1,q_2,p_1,p_2) &=& \frac{1}{2}(p_1^2 + p_2 ^2 + q_1^2 + q_2^2),
  \end{eqnarray}
  where the potential $V_1(q_1,q_2)$ is a homogeneous cubic or
  quartic function of $q_{1}$, $q_{2}$ and $\epsilon$ is a small 
  parameter, i.e. $|\epsilon|<<1$. 
  If renormalization group equation of the system
  (\ref{Hamiltonian}) is a Hamiltonian system
  up to the second leading order of $\epsilon$, then the original system
  (\ref{Hamiltonian}) and the renormalization group equation are
  integrable.}

\vspace*{1ex}
Let us briefly review renormalization group method by using a simple 
system with one degree of freedom:
\begin{equation}
	H(q,p) = H_0(q,p) + \epsilon V_1(q),
\end{equation}
\begin{equation}
	H_0(q,p) = \frac{1}{2}(p^{2}+q^{2}), \quad
	V_1(q) = \frac{1}{2}q^{2}.
\end{equation}
The equation of motion is
\begin{equation}
   \frac{\D^{2}q}{\D t^{2}}+q=-\epsilon q,	\label{eq:ex}
   \label{eq:sample}
\end{equation}
and the exact solution is 
\begin{equation}
   q= B_{0}\cos(\sqrt{1+\epsilon}t)+C_{0}\sin(\sqrt{1+\epsilon}t), 
\end{equation}
where $B_{0}$ and $C_{0}$ are constants of integration and determined 
by initial condition at the initial time $t=t_{0}$. We perturbatively 
solve Eq.(\ref{eq:ex}) by expanding  $q$ as a series of 
positive powers of $\epsilon$:
\begin{equation}
	q=q_{0}+\epsilon q_{1}+\epsilon^{2}q_{2}+\cdots.
\end{equation}
This naive expansion gives
\begin{eqnarray}
	& & \hspace*{-2em} q(t;t_0,B_0,C_0) 
	= B_{0}\cos t+C_{0}\sin t+\frac{\epsilon}{2}(t-t_{0})(C_{0}\cos 
	t-B_{0}\sin t) \nonumber \\
	&&\hspace*{-1em}-
	\frac{\epsilon^{2}}{8}\left[(t-t_{0})(C_{0}\cos t-B_{0}\sin 
	t)+(t-t_{0})^{2}(B_{0}\cos t+C_{0}\sin t)\right]+O(\epsilon^{3}),
\end{eqnarray}
and this expansion breaks 
when $\epsilon(t-t_{0})\geq 1$ because of secular terms. 
Here we ignored homogeneous parts of $q_{j} (j\ge 1)$ 
which is the kernel of the linear operator $L\equiv d^{2}/dt^{2}+1$, 
since they can be included in $q_{0}$. 
Renormalization group method removes the secular terms
by regarding $B_0$ and $C_0$ as functions of the initial time $t_0$,
and the evolutions of $B(t_0)$ and $C(t_0)$ are determined by 
RGE~\cite{chen-96,kunihiro-95}:
\begin{equation}
	\left.\frac{\partial q}{\partial 
	t_{0}}\right|_{t_{0}=t}=0,~~\hbox{for any}~t \label{eq:rg}.
\end{equation}
In other words, the RGE is
\begin{subeqnarray}
	 \label{eq:rgex}
	\frac{\D B(t)}{\D t}&=&\left(\frac{\epsilon}{2}
	-\frac{\epsilon^{2}}{8}\right)C(t)+O(\epsilon^{3}), \\
	\frac{\D C(t)}{\D t}&=&-\left(\frac{\epsilon}{2}
	-\frac{\epsilon^{2}}{8}\right)B(t)+O(\epsilon^{3}),
\end{subeqnarray}
and it gives the renormalized solution $q^{\mbox{\scriptsize RG}}$:
\begin{eqnarray}
	q^{\mbox{\scriptsize RG}}(t) 
	&=& q(t;t,B(t),C(t)) \nonumber \\
 	&=& B_{0}\cos\left(\left(1+\frac{\epsilon}{2}
	-\frac{\epsilon^{2}}{8}\right)t\right)+
	C_{0}\sin\left(\left(1+\frac{\epsilon}{2}
	-\frac{\epsilon^{2}}{8}\right)t\right)+O(\epsilon^{3}),
\end{eqnarray}
where $B_{0}=B(t_{0})$ and $C_{0}=C(t_{0})$. 
This expression gives an approximate but global solution to 
Eq.(\ref{eq:sample}) up to $O(\epsilon^{2})$. 

In the RGE (\ref{eq:rgex}) $B(t)$ and $C(t)$ are canonical 
conjugate variables because the equation is yielded by Hamiltonian
\begin{equation}
	\HRG=\left(\frac{\epsilon}{4}-\frac{\epsilon^{2}}{16}\right)
	(B^{2}+C^{2}),
\end{equation}
and
\begin{subeqnarray}
	\frac{\D B}{\D t}
	&=&\frac{\partial \HRG}{\partial C}+O(\epsilon^{3}),\\
	\frac{\D C}{\D t}
	&=&-\frac{\partial \HRG}{\partial B}+O(\epsilon^{3}).
\end{subeqnarray}
Moreover, the system governed by the $\HRG$ is obviously integrable 
and $(B^{2}+C^{2})/2$, which corresponds to $H_0$, 
is an integral of $\HRG$. 
Do these statements hold even in systems where chaotic motion appears? 

To answer this question,  we calculate RGEs in system~(\ref{Hamiltonian}),
whose equation of motion is
\begin{equation}
  \label{eq-motion}
  \frac{\D^2 q_j}{\D t^2} + q_j = -\epsilon\fracp{V_1}{q_j},
	\quad (j=1,2).
\end{equation}
First let potential $V_{1}$ be homogeneous quartic functions
of $q_{1}, q_{2}$:
\begin{equation}
   V_1(q_1,q_2) = \al1 q_1^4 + \al2 q_1^3 q_2 + \al3 q_1^2 q_2^2
  + \al4 q_1 q_2^3 + \al5 q_2^4 \label{eq:pot}.
\end{equation}
We expand $q_1$ and $q_2$ as
\begin{equation}
  \label{expand-2}
  q_j = q_{j,0} + \epsilon q_{j,1} + \epsilon^2 q_{j,2} + \cdots,
	\quad (j=1,2),
\end{equation}
and write the zero-th order solution to 
the equation~(\ref{eq-motion}) as
\begin{equation}
  \label{zero-sol}
  q_{j,0} = B_j \cos(t) + C_j \sin(t), \quad (j=1,2),
\end{equation}
where we omit subscript $0$ of $B_j$ and $C_j$ $\ (j=1,2)$ for
simplicity of symbols.
Following the  procedure to obtain RGE, we get 
\begin{subeqnarray}
  \label{renormal-eq}
  \frac{\D B_j}{\D t} = f_j(B_1,C_1,B_2,C_2; \epsilon,\al1,\cdots,\al5), \\
  \frac{\D C_j}{\D t} = g_j(B_1,C_1,B_2,C_2; \epsilon,\al1,\cdots,\al5),
\end{subeqnarray}
although the explicit forms of $f_j$ and $g_j$ are not shown here
because of complexity.

To clarify the condition of $\al1,\cdots,\al5$ with which
the RGEs (\ref{renormal-eq}) become Hamiltonian systems,
we define $\Delta_j$ as
\begin{equation}
  \Delta_j \equiv \fracp{f_j}{B_j} + \fracp{g_j}{C_j},
  \qquad (j=1,2).
\end{equation}
For a RGE, both of the equations $\Delta_1=0$ and $\Delta_2=0$ 
are satisfied if and only if Hamiltonian exists and 
\begin{subeqnarray}
  \label{renormal-eq-H}
  \frac{\D B_j}{\D t} &=& \fracp{\HRG}{C_j}, \\
  \frac{\D C_j}{\D t} &=& -\fracp{\HRG}{B_j}.
\end{subeqnarray}
We show the $\Delta_j$ up to the second order of $\epsilon$:
\begin{subeqnarray}
  \label{quadra-delta}
  \Delta_1 &=& \frac{\epsilon^2}{8} (B_1 C_2 - C_1 B_2)
  \left\{
    ( 9\al2^2 +4\al3^2 -24\al1\al3 -9\al2\al4 )(B_1 B_2 + C_1 C_2)
  \right. \nonumber\\
  &+& \left. 3 [ (\al2 + \al4) \al3
    - 6 (\al1\al4 +\al2\al5) ] (B_2^2 + C_2^2) \right\},\\
  \Delta_2 &=& -\frac{\epsilon^2}{8} (B_1 C_2 - C_1 B_2)
  \left\{
    ( 9\al4^2 +4\al3^2 -24\al3\al5 -9\al2\al4 )(B_1 B_2 + C_1 C_2)
  \right. \nonumber\\
  &+& \left. 3 [ (\al2 + \al4)\al3
    -6(\al1\al4 +\al2\al5) ] (B_1^2 + C_1^2)
    \right\}.
\end{subeqnarray}
The $\Delta_1$ and $\Delta_2$ take zeros irrespective of values of 
$B_j$ and $C_j$ 
if and only if the three equations hold:
\begin{subeqnarray}
  \label{quartic-condition}
  & & 9\al2^2 +4\al3^2 -24\al1\al3 -9\al2\al4 = 0, \\
  & & 9\al4^2 +4\al3^2 -24\al3\al5 -9\al2\al4 = 0, \\
  & & (\al2 + \al4)\al3 -6 (\al1\al4 +\al2\al5) =0.
\end{subeqnarray}
Consequently, RGEs are Hamiltonian systems
when the coefficients of $V_1$, namely $\al1,\cdots,\al5$, 
satisfy the condition (\ref{quartic-condition}).
We remark that $\Delta_j$ has no terms of order $\epsilon$,
which is the leading order of Eq.(\ref{renormal-eq}),
and hence the RGEs are Hamiltonian systems for any coefficients
up to $O(\epsilon)$.

To understand the meaning of the condition (\ref{quartic-condition}), 
we use the Bertrand-Darboux theorem\cite{hietarinta-87}. 
with which we judge integrability of ``natural'' Hamiltonian 
systems with two degrees of freedom whose form is
\begin{equation}
	H=\frac{1}{2}(p_{x}^{2}+p_{y}^{2})+V(x,y).
\end{equation}
The theorem states that the following three conditions are 
equivalent: 

(1) There is an integral which is independent of 
Hamiltonian and quadratic with respect to the momenta. 

(2) For a set of constants, $(a,b,b',c_{1},c_{2}) \neq 
(0,0,0,0,0)$, the potential $V(x,y)$ satisfies the Darboux equation
\begin{eqnarray}
	\label{Darboux-eq}
	&&(V_{yy}-V_{xx})(-2axy-b'y-bx+c_{1}) \nonumber \\
	&&+2V_{xy}(ay^{2}-ax^{2}+by-b'x+c_{2}) \nonumber \\
	&&+V_{x}(6ay+3b)+V_{y}(-6ax-3b')=0,
\end{eqnarray}
where $V_{x}=\partial V/\partial x, V_{xx}=\partial^{2}V/\partial 
x^{2}$ and so on. 

(3) The system is separable in Cartesian, polar, elliptic or parabolic coordinates.
\vspace{1ex} \\ 

We apply the Bertrand-Darboux theorem
to the original Hamiltonian systems written by Eqs.(\ref{Hamiltonian}) 
and (\ref{eq:pot}) 
to judge whether the systems are integrable or not.
The Darboux equation restricts the coefficients $\al1,\cdots,\al5$ in 
the potential function (\ref{eq:pot}) (Table I),
and we can find in the Table \ref{tab:quartic} 
the condition (\ref{quartic-condition}).
Consequently, when RGEs become Hamiltonian systems,
the original systems are integrable.
Moreover, the RGEs are also integrable
since they have another integral $I$,
\begin{equation}
	I\equiv (B_{1}^{2}+C_{1}^{2}+B_{2}^{2}+C_{2}^{2})/2,
\end{equation}
except for Hamiltonian $\HRG$ itself.
Note the quantity $I$ corresponds to $H_0$ in Eq.(\ref{Hamiltonian}).

\begin{table}[hbtp]
	\caption{Condition of $\al1,\cdots,\al5$,
	to satisfy the Darboux equation.
	Here $\al1,\cdots,\al5$ are the coefficients of 
	homogeneous quartic potential functions.
	There are two cases I and II,
	and case II must simultaneously satisfy
	the three equations, which coincide with the condition
	that RGEs become Hamiltonian systems.}
   \label{tab:quartic}
  \begin{center}
    \begin{tabular}{c|c|c}
        \hline \hline
                & Condition for $\alpha_j$
                & Potential function \\
        \hline
        I 
                & $\al2=\al4=0, ~\al3=2\al1=2\al5$
                & $V_1 = \al1 (q_1^2 + q_2^2)^2$ \\
        \hline
                &   $9\al2^2 +4\al3^2 -24\al1\al3 -9\al2\al4 = 0$   
                &  $V_1= \al1 q_1^4 + \al2 q_1^3 q_2 + \al3 q_1^2 
                    q_2^2$ \\
        II
		& $9\al4^2 +4\al3^2 -24\al3\al5 -9\al2\al4 = 0$
		& $ + \al4 q_1 q_2^3 + \al5 q_2^4 $\\
		&$(\al2 + \al4)\al3 -6 (\al1\al4 +\al2\al5) =0$
                & \\   
        \hline
    \end{tabular}
      \end{center}
\end{table}
\noindent 


For the potential of homogeneous cubic functions of $q_{1}$ and $q_{2}$,
\begin{equation}
	V_{1}(q)=\al1 q_{1}^{3}+\al2 q_{1}^{2}q_{2}+\al3 q_{1}q_{2}^{2}+\al4 
	q_{2}^{3},
\end{equation}
secular terms appear only in even orders of $\epsilon$ 
and the leading and the second leading orders of RGEs are 
$\epsilon^{2}$ and $\epsilon^{4}$, respectively. 
Then we calculate $\Delta_{1}$ and $\Delta_{2}$ up to the fourth order 
of $\epsilon$ and
\begin{subeqnarray}
	\Delta_{1}&=&\frac{\epsilon^{4}}{108}(B_{1}C_{2}-B_{2}C_{1})
	   (3(\al1\al3+\al2\al4)-\al2^{2}-\al3^{2})h_{1}, \\
	\Delta_{2}&=&-\frac{\epsilon^{4}}{108}(B_{1}C_{2}-B_{2}C_{1})
	   (3(\al1\al3+\al2\al4)-\al2^{2}-\al3^{2})h_{2},
\end{subeqnarray}
where
\begin{subeqnarray}
	h_{1}&=& 40\al2(3\al1+\al3)(B_{1}^{2}+C_{1}^{2}) \nonumber \\   
	&+& (66\al1\al2+139\al2\al3-495\al1\al4+186\al3\al4)
		(B_{2}^{2}+C_{2}^{2}) \nonumber \\
	&+& 6(33\al1^{2}+34\al2^{2}+12\al3^{2}-40\al1\al3+15\al2\al4)
		(B_{1}B_{2}+C_{1}C_{2}), \\
	h_{2}&=& 40\al3(3\al4+\al2)(B_{2}^{2}+C_{2}^{2}) \nonumber \\
	&+& (66\al3\al4+139\al2\al3-495\al1\al4+186\al1\al2)
		(B_{1}^{2}+C_{1}^{2}) \nonumber \\
	&+& 6(33\al4^2+34\al3^2+12\al2^2-40\al2\al4+15\al1\al3)
		(B_{1}B_{2}+C_{1}C_{2}).
\end{subeqnarray}
One condition for $\Delta_{1}=\Delta_{2}=0$ is $h_1=h_2=0$
whose unique solution is the trivial one,
i.e. $(\al1, \al2, \al3, \al4)=(0,0,0,0)$,
and hence we ignore this condition.
The other condition is $3(\al1\al3+\al2\al4)-\al2^{2}-\al3^{2}=0$,
and this condition is found in Table \ref{tab:cubic}
which shows the condition for the Darboux equation being satisfied.
Consequently, the original systems (\ref{Hamiltonian}) 
with the cubic perturbation are integrable 
when RGEs are Hamiltonian systems up to the second leading order. 
We therefore proved our Theorem.

\begin{table}[hbtp]
   \caption{Condition of $\al1,\cdots,\al4$,
	to satisfy the Darboux equation.
	Here $\al1,\cdots,\al4$ are the coefficients of 
	homogeneous cubic potential functions.
	Four cases I,$\cdots$, IV are obtained,
	and the condition in case II coincides to the one
	that RGEs become Hamiltonian systems.}
    \label{tab:cubic}
  \begin{center}
    \begin{tabular}{c|c|c}
        \hline \hline
                & Condition of $\alpha_j$
                & Potential function \\
        \hline
        I
                & $\al1=\al2=\al3=\al4=0$
                & $V_1 = 0$ \\
       II
                & $3(\al1\al3+\al2\al4)-\al2^{2}-\al3^{2}=0$
                & $V_{1}=\al1 q_{1}^{3}+\al2 q_{1}^{2}q_{2}$ \\
                & 
                & $+\al3 q_{1}q_{2}^{2}+\al4 q_{2}^{3}$ \\
       III
                & $\al1=2\al3, ~\al2=\al4=0$
                & $V_1 = \al3 (2 q_1^3 + q_1 q_2^2)$ \\
       IV
                & $\al1=\al3=0, ~2\al2=\al4$
                & $V_1 = \al2 (2 q_2^3 + q_1^2 q_2)$ \\
        \hline
    \end{tabular}
     \end{center}
\end{table}


In summary, we investigated dynamical properties of 
renormalization group equation (RGE) 
which are symplectic properties, integrability
and integrals.
Hamiltonian systems with two degrees of freedom were considered,
whose integrable part is harmonic oscillators
and homogeneous cubic and quartic potential functions are added 
as perturbation.
The main result of this article is that
the original Hamiltonian systems and RGEs are integrable
if the RGEs are Hamiltonian systems up to the second leading order
of a small parameter.

We have two future works.
One is generalization of potential function.
We can directly compare the Darboux equation
with condition for RGE being Hamiltonian systems
if we write RGEs for the general potential.
The other is extension to systems with higher degrees of freedom.
For the purpose, extended Bertrand-Darboux theorem \cite{marshall-88}
is available.
Taking the contraposition of our main result,
we suppose that we can obtain information on chaotic
behavior by considering how symplectic properties of RGEs
break in non-integrable systems.

We express our thanks to
Atsushi Taruya and Tetsuro Konishi
for valuable discussions.
One of the authors Y.Y.Y. acknowledges
helpful suggestions and discussions to
Masaharu Ishii.
Y.Y.Y. also thanks
Kazuhiro Nozaki
and members of R-lab. at Nagoya university for fruitful comments. 
Y.N. is  supported in part by the Grand-In-Aid 
for Scientific Research of the Ministry of Education, Sports and 
Culture of Japan(09740196).


\end{document}